# Linear Viscoelasticity of Soft Glassy Materials

Manish Kaushal, and Yogesh M. Joshi*

Department of Chemical Engineering,

Indian Institute of Technology Kanpur, Kanpur 208016 INDIA

* Corresponding Author, E- mail: joshi@iitk.ac.in,
Tel.: +91-512-2597993, Fax: +91-512-2590104

Abstract

Owing to lack of time translational invariance, aging soft glassy materials do not obey fundamental principles of linear viscoelasticity. We show that by transforming the linear viscoelastic framework from the real time domain to the effective time domain, wherein the material clock is readjusted to account for evolution of relaxation time, the soft glassy materials obey effective time translational invariance. Consequently, we demonstrate successful validation of principles of linear viscoelasticity (Boltzmann superposition principle and convolution relation for creep compliance and stress relaxation modulus) for different types of soft glassy materials in the effective time domain.

Knowledge of rheological constitutive equation that relates stress to strain and its time derivatives is essential to understand the flow behavior of any material. Important class of constitutive equations pertains to equilibrium soft materials that usually follow fundamental symmetry laws of nature such as time translational invariance (TTI). In constitutive equations that follow TTI replacement of time variable $t$ by $t+a$, where $a$ is a real number, does not change the nature of the constitutive equation.[1] Glassy soft materials such as concentrated emulsions and suspensions, pastes, colloidal gels, foam etc., on the other hand, are out of thermodynamic equilibrium and do not follow TTI.[2] Their natural tendency to explore the phase space leads to slow reorganization of microstructure as a function of time, the process typically known as physical aging. Application of deformation field slows down the reorganization or breaks the structure thereby demonstrating yield stress[3] and thixotropy.[4] Consequently glassy soft materials do not obey the fundamental principles of linear viscoelasticity including Boltzmann superposition principle (BSP).[5] Owing to vast academic interest and industrial applications this class of materials has attracted significant attention of the scientific community over the past decade.[6-9] However inapplicability of fundamental laws of viscoelasticity along with yielding behavior and thixotropy limit the analyzability and optimal applicability of these materials. Recently Joshi and coworkers[10-12] showed that fundamental laws of linear viscoelasticity such as Boltzmann superposition principle and time temperature superposition can be applied to aging soft glassy materials by transforming the same from the real time domain to the effective time domain, wherein material clock is readjusted to obliterate time dependency. In this work we take the validation of linear viscoelasticity to the next level. By carrying out experiments on various soft glassy materials with different microstructures, we show that aging glassy materials validate the convolution principle relating the various response functions in the effective time domain.

Ergodic materials that demonstrate TTI also obey BSP.[13] Depending upon whether the deformation field is stress ($\sigma$) controlled or strain ($\gamma$) controlled, BSP can be written in form: $\gamma(t) = \int_{-\infty}^{t} J(t - t_w) \dot{\sigma} dt_w$ or $\sigma(t) = \int_{-\infty}^{t} G(t - t_w) \dot{\gamma} dt_w$, where $J$ is creep compliance, $G$ is stress relaxation modulus, $t$ is the present time while $t_w$ is the past time at which deformation field was applied. The time derivatives $\dot{\sigma}$ and $\dot{\gamma}$ are with respect to the aging time $t_w$. The important characteristic feature of applicability of TTI is that $J$ as well as $G$ are only functions of time elapsed since application of deformation field: $J = J(t - t_w)$ and $G = G(t - t_w)$.[13] Furthermore, equating Laplace transform of both the expressions of BSP lead to convolution relation between creep compliance and stress relaxation modulus given by: $t = \int_0^t G(s) J(t-s) ds$.[13] This relation is frequently used in the rheology literature as it gives complete information about the rheological behavior of a material with the knowledge of any response function.

In the aging glassy materials, on the other hand, owing to time dependent physical properties, TTI is not applicable.[5] Compliance and relaxation modulus, therefore, show additional dependence on the time at which deformation field was applied: $J = J(t - t_w, t_w)$ and $G = G(t - t_w, t_w)$. Consequently neither BSP, nor the convolution relation is applicable to the aging glassy materials. We investigate this behavior for three different types of soft glassy materials namely, aqueous suspension of Laponite RD, acrylic emulsion paint and polymer – cloisite clay nanocomposite. All the three materials demonstrate physical aging wherein they undergo time dependent evolution of microstructure. Laponite suspension constitutes disk-like clay particles dispersed in water having dissimilar charges. The slow evolution of their microstructure is due to exploration of the phase space to attain progressively lower free energy states in a very complex energy landscape attributed to repulsive and attractive interactions among the particles.[14] In this work we use 3.5 weight % of suspension Laponite RD (Southern Clay Products Inc.). The suspension was prepared by dispersing oven dried white powder of Laponite RD in ultrapure water under vigorous stirring. The freshly prepared suspension was stored in polypropelene bottles for 2 months before carrying out the experiments. The detailed procedure of preparing Laponite suspension can be found elsewhere.[15] The second system is commercially procured acrylic emulsion paint (solid content 48% ± 0.03% w/w, Kansai Nerolac Paints Limited, Mumbai) is a concentrated emulsion with water as a continuous phase, and is used as it is. Owing to high concentration, acrylic drops acquire nonspherical structure. Aging in such system involves evolution of structure to lower the interfacial surface area as a function of time.[16] The third system used in this work is polymer - clay nanocomposite composed of epoxy polymer blended with cloisite 10A®. Cloisite 10A is obtained by organically modifying bentonite clay with benzyl (hydrogenated tallow alkyl) dimethyl group to render it affinity with epoxy. In a shear melting process, sheet-like clay particles of Cloisite get oriented in the gradient direction. Physical aging in polymer clay nanocomposite involves disorientation dynamics (randomization) of clay plates subsequent to shear melting.[17] The nanocomposite was prepared by blending Epoxy LY 556 (Dow chemicals) with Cloisite 10A (Southern clay products) using mechanical stirrer for 10 min followed by 4 h of sonication.

The rheological experiments were conducted using stress controlled rheometer (Anton Paar MCR 501). For Laponite gel and emulsion paint we employed concentric cylinder geometry (inside diameter of 5 mm and gap of 0.2 mm), while for Polymer – clay nanocomposite we used cone and plate geometry with 25 mm diameter and angle 1°. Before every experiment the samples were shear melted under the large amplitude oscillatory strain ($6\times10^3$ % for Laponite suspension, $3\times10^4$ % for emulsion paint, and $5\times10^5$ % for nanocomposite at frequency 0.1Hz) to erase the deformation/aging history, and allowed to age for a predetermined time (aging time, $t_w$). In this work we perform two types of experiments: creep and stress relaxation at various aging times subsequent to shear melting. In the creep experiments step stress (0.3 Pa for Laponite suspension, 2 Pa for emulsion paint, and 25 Pa for nanocomposite) was applied and corresponding evolution of compliance was measured. On the other hand, in stress relaxation experiments step strain (0.5 % for Laponite suspension, 1 % for emulsion paint, and 5 % for nanocomposite) was applied to the samples and subsequent relaxation of stress was measured. We verified that the applied magnitudes of creep stress and step strain are in the linear viscoelastic domain.

In figure 1(a) we plot compliance induced in the material following application of creep stress, while in figure 1(b) we plot relaxation modulus subsequent to application of step strain at different times elapsed since stopping the shear melting for Laponite suspensions. The same data for the other two materials is shown in supplementary information figures S1 and S2. It can be seen that compliance induced in the material is smaller and relaxation of stress is slower for experiments carried out at greater aging time $t_w$. This confirms breakdown of TTI, wherein $J$ and $G$ do not depend only on $t - t_w$ but show additional dependence on $t_w$. In addition, owing to aging during the course of experiments, the curvature of the creep/stress relaxation curves is not self-similar in order to obtain superposition by merely horizontal shifting. For materials whose properties change with time, irrespective of whether spontaneously or externally by changing the temperature, Hoppkins[18] proposed an effective time scale given by: $\xi(t) = \tau_0 \int_0^t dt'/\tau(t')$. Consequently transformation of time variable from the real time with time dependent relaxation time $\tau(t')$ to the effective time with constant relaxation time $\tau_0$ leads to readjustment of the material clock so as to eliminate the time dependency. Such transformation implies that the relaxation that occurs over time $t$ with time dependent relaxation time $\tau(t')$ in the real time domain is equivalent to what occurs in the effective time domain over time $\xi$ with constant relaxation time $\tau_0$. Boltzmann superposition can then be expressed in effective time domain as:[10, 19]

$$\sigma(\xi) = \int_{-\infty}^{\xi} G(\xi - \xi_w) \frac{d\gamma}{d\xi_w} d\xi_w, \text{ and} \tag{1a}$$

$$\gamma(\xi) = \int_{-\infty}^{\xi} J(\xi - \xi_w) \frac{d\sigma}{d\xi_w} d\xi_w, \tag{1b}$$

where $\xi_w = \xi(t_w)$, effective time at which deformation field was applied. In the effective time domain compliance and relaxation modulus depend on effective time elapsed since application of deformation field: $\xi - \xi_w$. However, in principle, in order to represent the effective time in terms of the real time, functional form of the dependence of relaxation time on the real time is necessary. Relaxation time of aging glassy materials is known to show power law dependence on real time given by:[5, 10, 19, 20] $\tau(t') = A\tau_m^{1-\mu} t'^{\mu}$, where $A$ is constant coefficient, $\tau_m$ is microscopic relaxation time and $\mu$ is power law coefficient. For this dependence, effective time elapsed since application of deformation field takes a form:

$$\xi - \xi_w = \tau_0 \int_{t_w}^{t} \frac{dt'}{\tau(t')} = \frac{\tau_0^{\mu}}{A} \left[ \frac{t^{1-\mu} - t_w^{1-\mu}}{1-\mu} \right]. \tag{2}$$

In equation (2) we assume $\tau_0 = \tau_m$. In figure 2 we plot compliance (a) and relaxation modulus (b) of Laponite suspension as a function of $\xi - \xi_w$ multiplied by a constant factor $A/\tau_0^{\mu}$. We also perform minor vertical shifting with vertical shift factors very close to unity (mentioned in the supporting information in figure S3). Usually the vertical shift factors are necessary when modulus also shows enhancement as a function of aging time. However when modulus does not show appreciable enhancement, vertical shift factors may still be required to account for minor vertical adjustment of the experimental data due to uncertainties. It can be seen that both compliance and relaxation modulus show exclusive dependence on $\xi - \xi_w$ leading to superposition for a value of $\mu$ =1.21±0.02. The value of $\mu$ can also be independently obtained by fitting suitable functional form such as KWW to the stress relaxation data. However owing to aging during the course of stress relaxation, only the short time relaxation data (in the limit of $t - t_w << t_w$) is suitable for fitting. As expected, fit of KWW function $G(t) = G_0 \exp\left[-(t/\tau)^\beta\right]$ to the short term stress relaxation data led to practically identical value of $\mu$ =1.24±0.002 with $\beta$ =0.285±0.014.

Existence of superposition of $J$ and $G$ confirms applicability of BSP in the effective time domain. This also suggests that in the effective time domain glassy materials validate *effective time translational invariance*. Taking Laplace transform of equations (1a) and (1b) leads to $\tilde{J}(p)\tilde{G}(p) = 1/p^2$ thereby resulting in convolution relation in the effective time domain:

$$\xi = \int_0^\xi G(\zeta) J(\xi - \zeta) d\zeta \tag{3}$$

Although the mathematical form of this expression is identical to the conventional form, equation (3) is applicable for aging glassy materials only in the effective time domain. Equation (3) is a Volterra integral of first kind, where information of either of $J$ or $G$ can be used to get the other. However obtaining solution of this kind of integral using either by Laplace transform or by algebraic method is an ill posed problem as small noise in $J$ or $G$ data propagates and leads to large errors in computation.[21, 22] We adopt a numerical scheme suggested by Zhu et al.,[21] wherein equation (3) is first differentiated using Leibnitz formula and then discretized the same to represent it in a matrix form. The prediction of $J(\xi - \xi_w)$ using experimental $G(\xi - \xi_w)$ is shown in figure 2(a), while that of $G(\xi - \xi_w)$ using experimental $J(\xi - \xi_w)$ is shown in figure 2(b). The predictions shown by the solid lines very nicely overlap with time-aging time superimposed data of $G$ as well as $J$, confirming validation of convolution relation in the effective time domain.

We apply similar procedure to the creep and relaxation modulus data of acrylic emulsion paint and polymer – clay nanocomposite. The complete data is shown in supplementary information figure S1 and S2. In figure 3, transformed data is plotted in the effective time domain (the corresponding vertical shift factors are shown in supplementary information figure S3), wherein creep and stress relaxation modulus data demonstrate excellent superposition. This suggests respective response functions depend only on the effective time elapsed since application of deformation field $(\xi - \xi_w)$ thereby validating BSP for very different kinds of soft glassy materials. We further use the convolution integral in the effective time domain to obtain $G$ (from experimental $J$) and $J$ (from experimental $G$). Similar to that shown in figure 2, solution of the convolution integral shown by thick line leads to an excellent prediction of the superposition in the effective time domain.

The predictions shown in figures 2 and 3 can be transformed from the effective time domain to the real time domain by inverting equation (2) as:

$$t - t_w = \left[\left\{A\tau_0^{-\mu}\left(\xi - \xi_w\right)\right\}(1-\mu) + t_w^{1-\mu}\right]^{1/(1-\mu)} - t_w, \quad (4)$$

where the term in braces is the abscissa of figures 2 and 3. The corresponding predictions of the creep compliance and the stress relaxation modulus is plotted in the real time domain for Laponite suspension in figure 1, while that of for emulsion and nanocomposite in figures S1 and S2. It can be seen that equation (4) leads to an excellent prediction of the experimental data.

Presently various theoretical approaches are available in the literature to model flow behavior of glassy soft materials. There are empirical 'toy' models that convey the qualitative physical behavior but are far from rigorous, and have limitation to be used as constitutive relation.[23] The other approaches such as mesoscopic soft glassy rheology model[5] and microscopic mode coupling theories,[24] while mathematically involved, present the rigorous constitutive relations. The use of Boltzmann superposition principle, on the other hand, is the most fundamental approach as it originates from the symmetry laws of nature. The present work facilitates application of well-established theories of linear viscoelasticity, including the Boltzmann superposition principle, to the glassy materials. Furthermore, linear viscoelastic framework presented here is not limited to only step strain or stress, but can be applied to any arbitrary flow fields which preserve the shape of the spectrum of relaxation times.

The principle finding of the present letter is validation of effective time – translational invariance (eTTI) when the material clock is adjusted to account for time dependent relaxation time, which allows successful estimation of one response function from the knowledge of other in the effective time domain. Physical origin of this behavior lies in the concept of effective time. The way effective time is defined it stretches the real time so as to keep the relaxation time in the effective time domain constant. This eliminates the time dependency; and consequently TTI gets validated in the effective time domain.

It is important to note that this is the first report wherein any response function is estimated from the other for any kind of time dependent materials in general and for any glassy material in particular. We feel that this is an extraordinary result which confirms usefulness of linear viscoelastic principles and opens up new frontiers of modeling approaches for soft glassy materials. The methodology described in this work can be extended to other types of glassy materials as well such as spin glasses and molecular glasses for different response functions beyond rheology. In addition, with the knowledge of appropriate dependence of relaxation time on time, this methodology can also be extended to chemical reactions wherein properties of the materials change very rapidly as a function of time.

To conclude we express fundamental principles of linear viscoelasticity such as Boltzmann superposition principle and convolution relation between creep compliance and relaxation modulus in the effective time domain by adjusting the material clock to eliminate the time dependency. For three soft glassy materials with different microstructure, we demonstrate validity of the mentioned principles of linear viscoelasticity by successfully predicting either of the response functions from the other. This result also confirms to applicability of effective time translational invariance in the effective time domain. We believe that this methodology can be applied to different kinds of time dependent processes let alone the aging phenomenon in glassy materials, and will open up new frontiers of modeling approaches.

**Acknowledgement:** We acknowledge financial support from the department of atomic energy – science research council (DAE-SRC), Government of India.

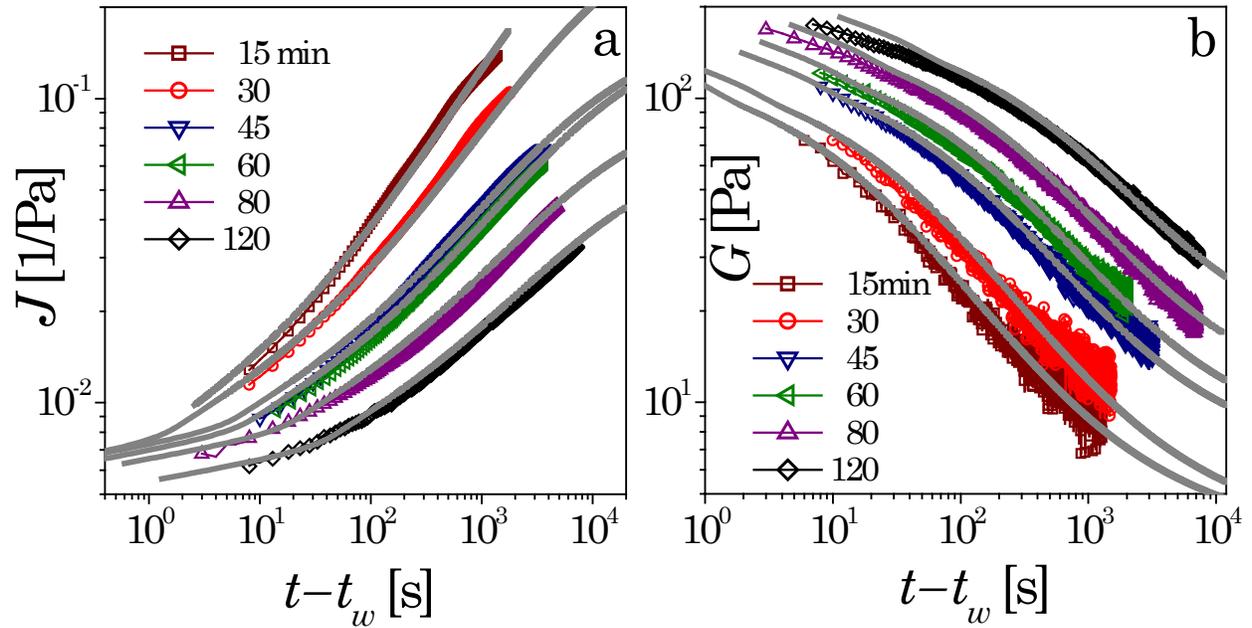

**Figure 1**: Compliance (a), and stress relaxation modulus (b) associated with Laponite suspension is plotted as a function of time elapsed since application of step stress and step strain respectively at different $t_w$. Solid lines in (a) are the predictions of compliance obtained from stress relaxation modulus shown in (b), while solid lines in (b) are the predictions of stress relaxation modulus obtained from compliance data shown in (a) using equations (3) and (4).

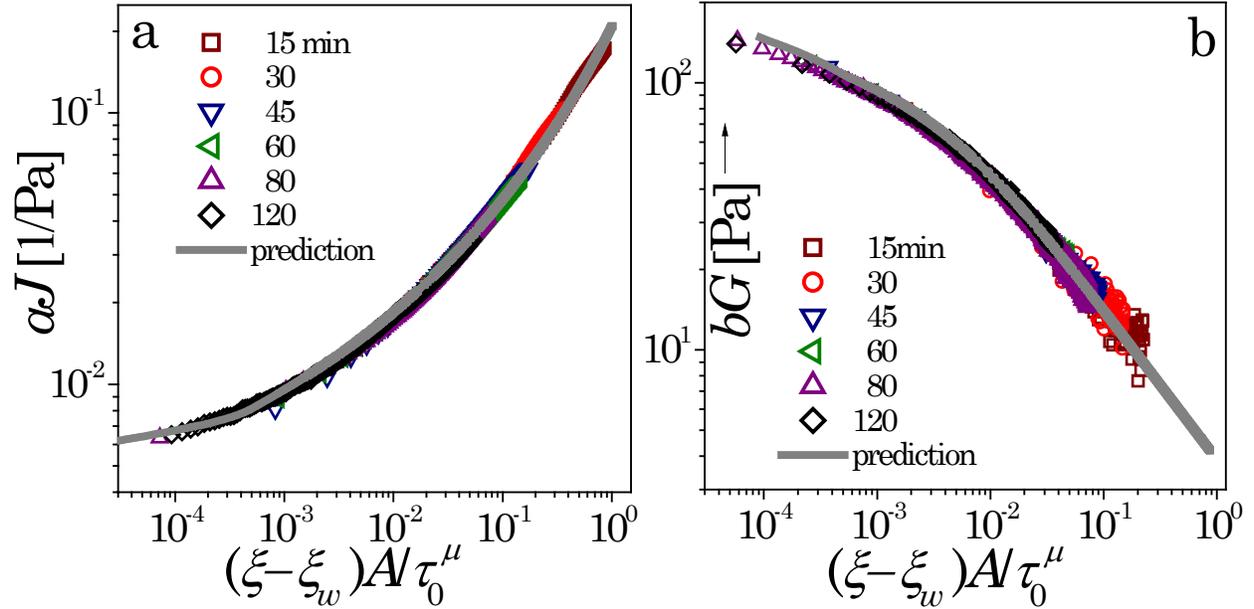

**Figure 2**: Superposition of compliance (a) and stress relaxation (b) in the effective time domain for $\mu=1.21\pm0.02$ for the data shown in figure 1. Solid line shown in (a) is prediction of compliance obtained from of stress relaxation data shown in (b) in the effective time domain, while solid line shown in (b) is prediction of stress relaxation modulus from compliance data shown in (a) by using convolution relation equation (3) in the effective time domain.

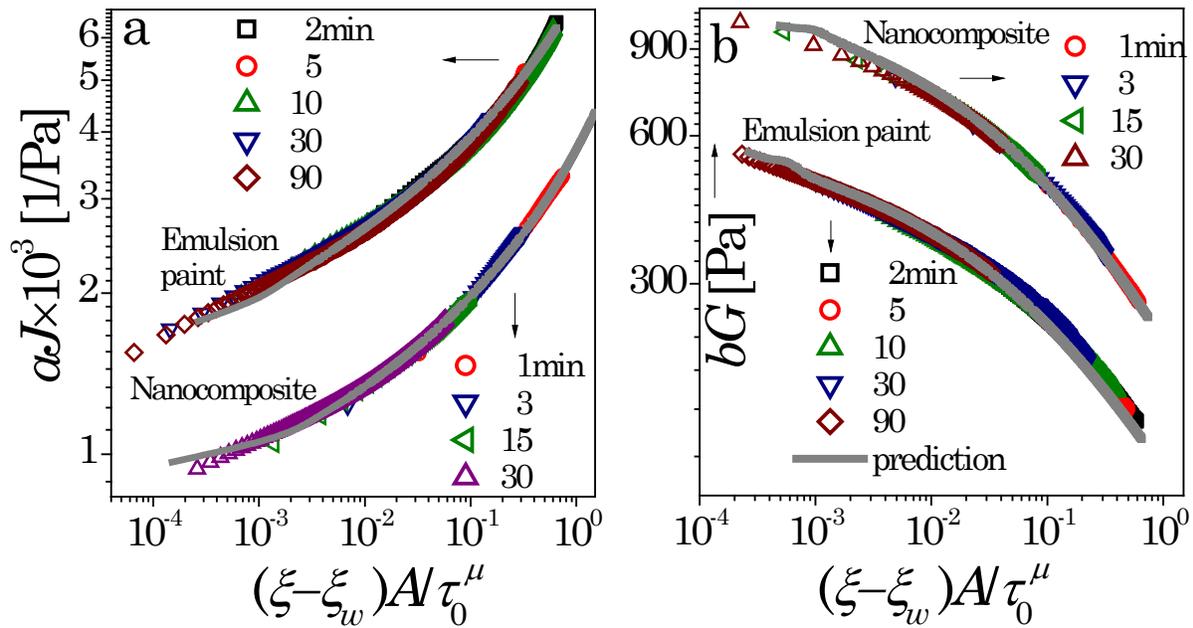

**Figure 3.** Superposition of compliance (a) and stress relaxation modulus (b) in the effective time domain for emulsion paint and polymer-clay nanocomposite for $\mu=1.18\pm0.01$ and $\mu=1.32\pm0.01$ respectively. Solid lines shown in (a) and (b) are the predictions of creep compliance and stress relaxation modulus respectively in the effective time domain using the convolution relation equation (3) in the effective time domain. Fitting KWW function to short time stress relaxation data leads to independent estimation of $\mu=1.19\pm0.001$ with $\beta=0.141\pm0.003$ for emulsion paint, and $\mu=1.30\pm0.002$ with $\beta=0.105\pm0.008$ for nanocomposite.

**Supplementary information:**

**Linear Viscoelasticity of Soft Glassy Materials**

Manish Kaushal, Yogesh M. Joshi*

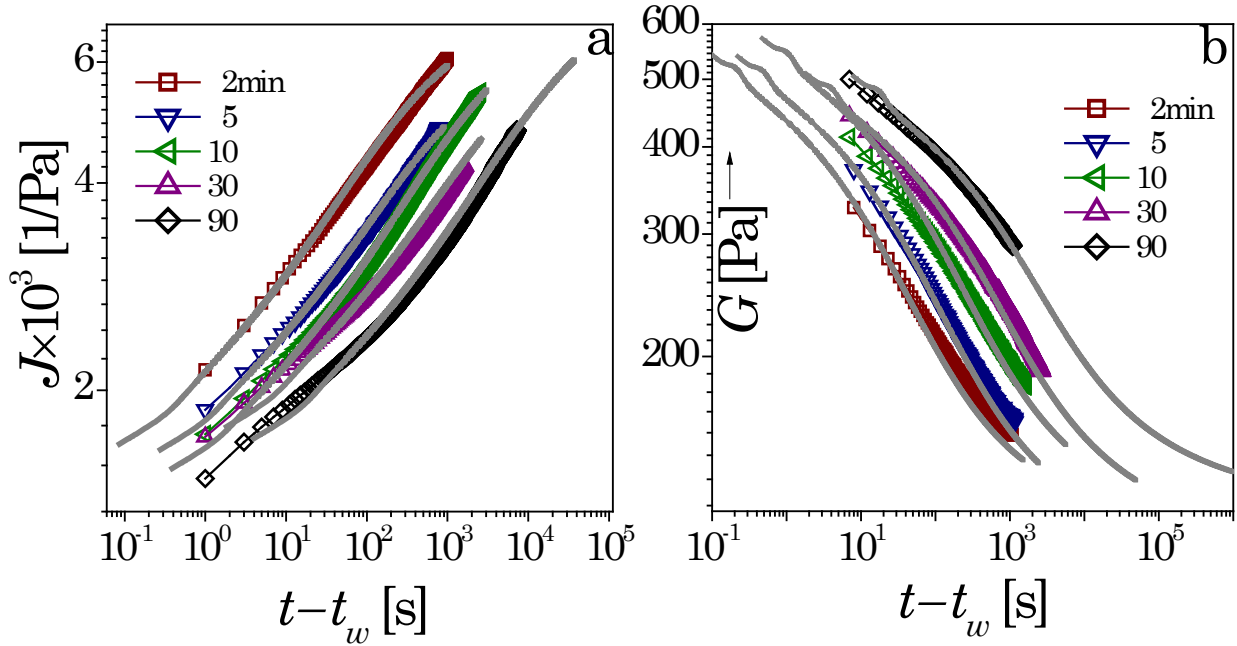

**Figure S1**: Compliance (a), and stress relaxation modulus (b) for Emulsion paint is plotted as a function of time subsequent to application of step stress and step strain respectively at different $t_w$. Solid lines in (a) show the predictions of compliance data obtained from stress relaxation modulus data shown in (b), and solid lines in (b) show the predictions of stress relaxation modulus obtained from compliance data shown in (a) using equations (3) and (4).

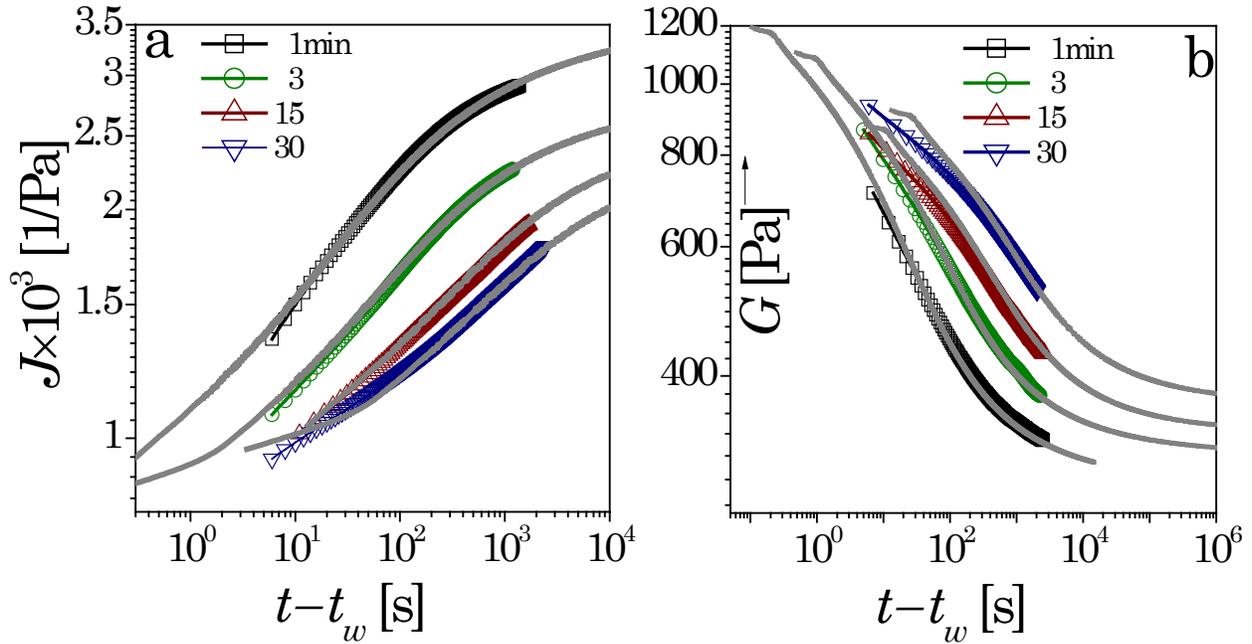

**Figure S2**: Compliance (a), and stress relaxation modulus (b) associated with Clay-nanocomposite is plotted as a function of time subsequent to application of step stress and step strain respectively at different $t_w$. Solid lines in (a) are the predictions of compliance data obtained from stress relaxation modulus data shown in (b), and solid lines in (b) are the predictions of stress relaxation modulus obtained from compliance data shown in (a) using equations (3) and (4).

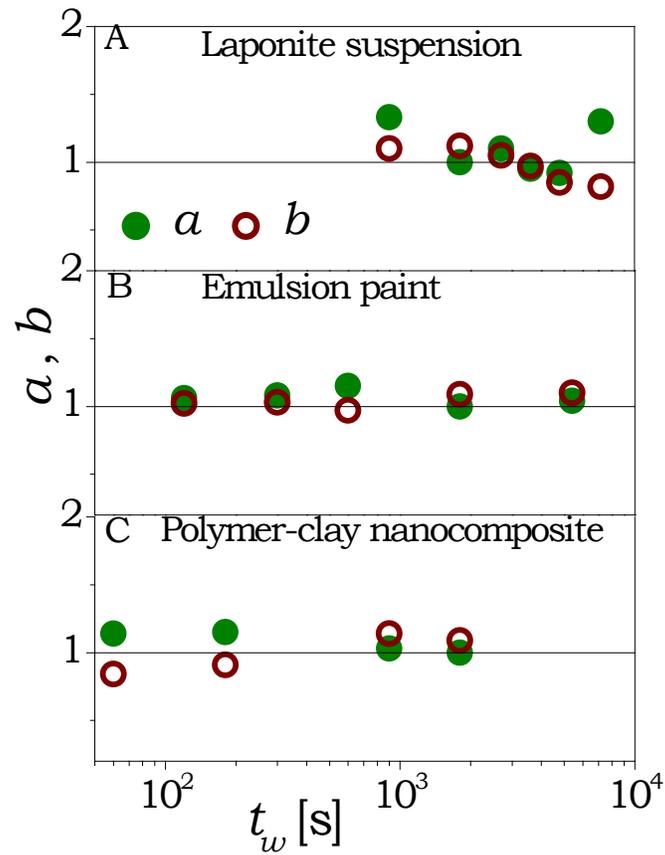

**Figure S3**: Vertical shift factors $a$ (for creep compliance, closed circles) and $b$ (for stress relaxation modulus, open circles) to get superposition shown in figures 2 and 3 are plotted as a function of waiting time $t_w$ for Laponite suspension (A), Emulsion paint (B), and Polymer-clay nanocomposite (C).